# Self-induced spin-orbit torque switching in a synthetic antiferromagnetic $Co_2MnGa$/MnGa bilayer

Rongkun Han[1,2*], Haohang Cheng[1], Dahai Wei[1], Yuan Lu[2], and Jianhua Zhao[1,3*]

[1]*State Key Laboratory of Semiconductor Physics and Chip Technologies, Institute of Semiconductors, Chinese Academy of Sciences, Beijing 100083, China*

[2]*Université de Lorraine, CNRS, Institut Jean Lamour, F-54011 Nancy, France*

[3]*State Key Laboratory of Spintronics, Hangzhou International Innovation Institute, Beihang University, Hangzhou 311115, China*

The large intrinsic spin current in magnetic Weyl semimetals (WSMs) provides a promising platform for spin-orbit torque (SOT) devices. Here, we demonstrate SOT-driven magnetization switching in a synthetic antiferromagnet (SAF) composed of a $Co_2MnGa$(CMG)/MnGa bilayer. In this heavy metal-free structure, CMG film functions simultaneously as a spin current source and as a magnetic layer. Macrospin simulations confirm the self-induced switching mechanism in the bilayer. The SOT generated by CMG switches the MnGa layer, which in turn triggers the reversal of the CMG layer via strong antiferromagnetic exchange coupling. These results elucidate the spin dynamics in SAFs and provide a new pathway toward SOT devices with self-induced magnetization switching.

[*]Corresponding authors: jianhuazhao@buaa.edu.cn and rongkun.han@univ-lorraine.fr

## Introduction

The spin Hall effect (SHE) and its inverse (ISHE) enable interconversion between charge and spin currents, providing a route for the electrical generation, manipulation, and detection of spin signals [1,2]. SHE-generated spin currents ($\boldsymbol{J}_s$) can exert spin-orbit torques (SOT) on an adjacent ferromagnet and drive magnetization precession or switching [3,4]. Materials with a large spin Hall angle ($\theta_{SH}$) are crucial for improving device performance and reducing power consumption in SOT devices [5-7]. Beyond conventional heavy metals with strong spin-orbit coupling (SOC), recent experimental and theoretical studies have shown that topological materials can exhibit a large intrinsic $\theta_{SH}$ [8-10], due to the Berry curvature in momentum space [11]. In Weyl semimetals (WSMs), Weyl nodes act as sources or sinks of Berry curvature, endowing these materials with unique topological and transport properties [12-15]. Co-based Heusler alloys such as $Co_2MnGa$ (CMG) and $Co_2MnAl$ (CMA) are WSMs, whose band structures produce a sizeable intrinsic SHE. Moreover, these materials have been shown to enable SOT-driven magnetization switching in adjacent ferromagnetic (FM) layers [16-18].

In addition to their topological properties, Co-based Heusler alloys possess favorable magnetic characteristics, including high Curie temperatures, low Gilbert damping, and high spin polarization, making them attractive candidates for spintronics devices [19,20]. The crystal structure of CMG includes $L2_1$, B2, $DO_3$, and A2 phases, depending on the degree of atomic ordering. The fully ordered $L2_1$ structure belongs to the space group *Fm-3m*, as illustrated in Fig. 1(a). The CMG films typically show in-plane magnetic anisotropy (IMA) due to the demagnetization field. However, perpendicular magnetic anisotropy (PMA) is desirable for low-power, high-speed, and high-density spintronic devices [21,22]. A synthetic antiferromagnet (SAF) structure, formed by coupling Co-based Heusler alloys with PMA MnGa films, provides a route to engineer PMA in the Heusler layer [23,24]. Such structures combine the advantages of low saturation magnetization, strong immunity to magnetic interference, and the ultrafast spin dynamics inherent to antiferromagnetic (AFM) materials [25-27].

Moreover, they can be effectively manipulated by external magnetic fields or spin currents and can readily be detected via the anomalous Hall effect (AHE) or tunneling magnetoresistance [28,29]. SAFs provides a promising solution to the challenge of controlling and detecting the Néel vectors in AFM, while maintaining the intrinsic advantages of AFM systems.

Here, we report the growth of high-quality $L2_1$-CMG films and their application in SOT-driven magnetization switching. We first demonstrate that the CMG acts as an efficient spin current source to switch the MnGa layer in a decoupled CMG/Al/MnGa multilayer. More importantly, self-induced magnetization switching is observed in an SAF CMG/MnGa bilayer. Macrospin simulations show that the spin current generated by CMG switches the PMA MnGa layer, which subsequently triggers the reversal of the CMG layer via strong AFM exchange coupling. These results provide a mechanism for self-induced switching in SAFs.

**Structure and magnetic properties**

Two types of multilayers, CMG(5 nm)/Al(2 nm)/MnGa(3 nm)/Al(3 nm) and CMG(1.5 nm)/MnGa(3 nm)/Al(3 nm) were grown on GaAs (001) substrates using molecular-beam epitaxy (MBE). The CMG layer was deposited at 350°C to obtain high crystalline quality. For the trilayer structure, a 2-nm-thick Al spacer layer was inserted to suppress atomic interdiffusion and magnetic coupling between CMG and MnGa. The MnGa layer was subsequently grown at 280°C, and a 3-nm-thick Al capping layer was deposited to prevent oxidation. The crystal structure and epitaxial relationship of the CMG/Al/MnGa multilayer were characterized by X-ray diffraction (XRD). As shown in Fig. 1(b), diffraction peaks corresponding to CMG (002) and (004) appear at 30.58° and 62.91°, giving a lattice constant of $c$=5.84 Å. In addition, a MnGa (002) peak at 52.29° confirms the formation of the $L1_0$ phase with $c$=3.50 Å. To determine the ordering of the CMG layer, we perform $\varphi$ scans on the {220} and {111} planes of CMG film, as well as the {220} plane of the GaAs substrate. As shown in Fig. 1(c), the {111} reflection exhibits a distinct fourfold symmetry, confirming that the CMG layer has the highest-order $L2_1$ structure [30]. Further, the epitaxial relationship is determined as CMG(001)[110]//GaAs(001)[110].

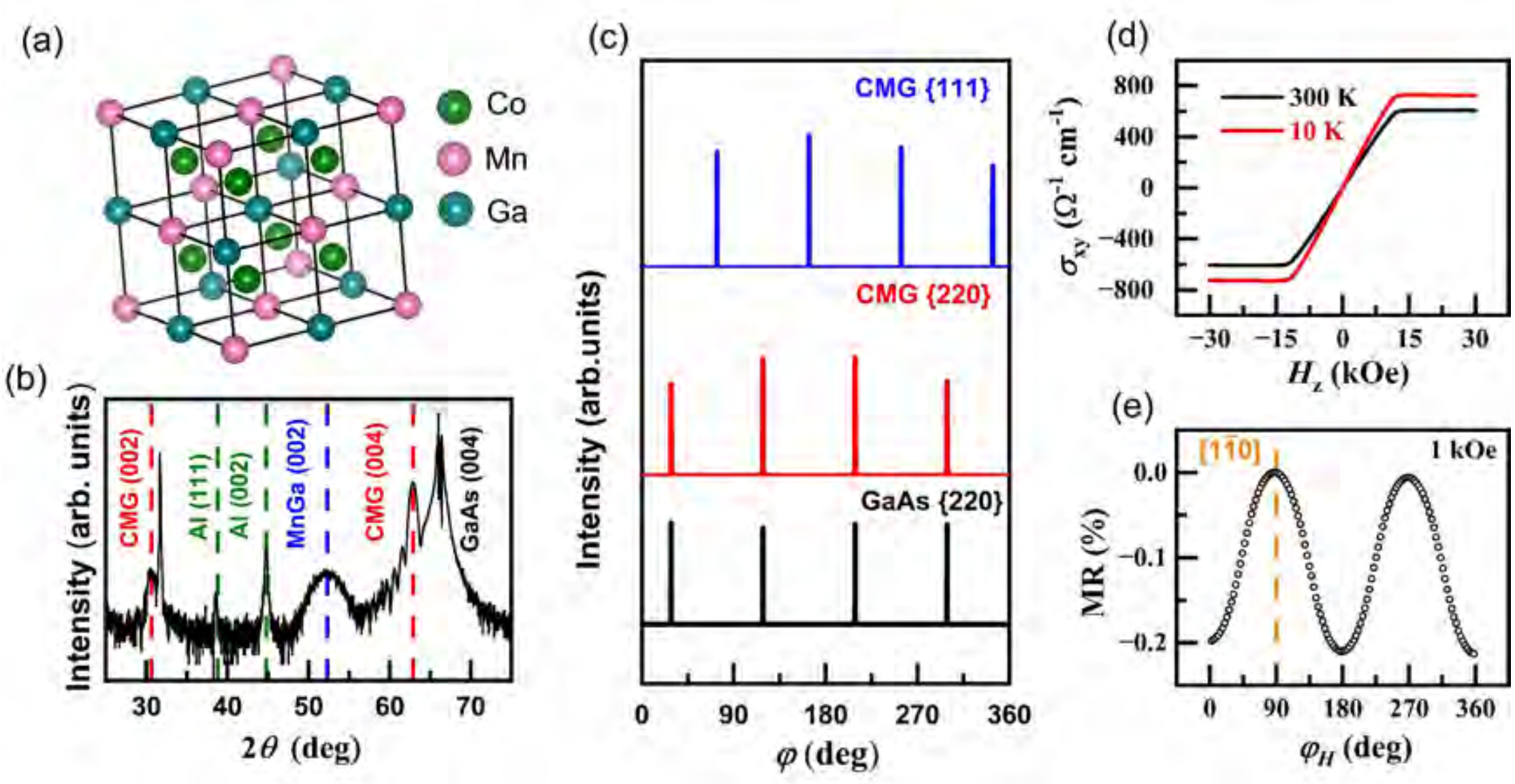

FIG. 1. (a) Crystal structure schematic of the $L2_1$-CMG Heusler alloy. (b) XRD $\theta$-2$\theta$ scan of the CMG/Al/MnGa multilayer. (c) XRD $\varphi$ scan of CMG {111} planes, CMG {220} planes, and GaAs {220} planes. (d) AHE curve of a CMG single layer at 300 K and 10 K. (e) AMR curves of CMG single layer at 1 kOe. The dash line indicates the [1-10] crystalline orientation.

To explore the magnetic and topological properties of the CMG film, a 30-nm-thick CMG single-layer film was also grown on GaAs substrate. The anomalous Hall conductivity (AHC) $\sigma_{xy}$ of the CMG film reaches 608 $\Omega^{-1}$ cm$^{-1}$ at 300 K and increases to 727 $\Omega^{-1}$ cm$^{-1}$ at 10 K, as shown in Fig. 1(d). This large AHC originates from the nontrivial band structure associated with a significant Berry curvature (Supplemental Material S1) [30,31]. Anisotropic magnetoresistance (AMR) measurements reveal a negative AMR with $R_\perp > R_{//}$ [Fig. 1(e)]. Here, $R_{//}$ and $R_\perp$ denote the magnetoresistances (MR) signals measured when the current is applied parallel and perpendicular to the in-plane magnetic field, respectively. Unlike conventional ferromagnets, which typically exhibit $R_{//} > R_\perp$ owing to stronger *s*–*d* scattering in the parallel configuration, CMG shows the opposite trend. This negative AMR indicates that charge transport is dominated by a single spin channel, implying a high degree of spin polarization characteristic of half-metallic systems [32-34]. Moreover, a negative MR is also observed in the CMG single layer (Supplemental Material S2) [31]. These results collectively suggest the magnetic WSM nature of our high-quality CMG film.

**Magnetization switching in decoupled system**

Figure 2(b) presents the in-plane and out-of-plane hysteresis loops of the CMG(5 nm)/Al(2 nm)/MnGa(3 nm) multilayer at 300 K. When the magnetic field is applied along the [110] or [1-10] crystalline orientations, the CMG layer magnetization rapidly saturates due to its IMA, with a saturation magnetization of $M_S$=762 emu/cm$^3$. The out-of-plane hysteresis reflects the combined response of the MnGa and CMG layers. At high fields (±13 kOe), the CMG magnetic moments are pulled out of the plane and reach saturation. At lower fields (±1.8 kOe), a distinct magnetization reversal of the MnGa layer is observed, indicating its strong PMA. The absence of magnetic coupling between the two layers identifies this structure as a decoupled system. Moreover, the $M_S$ of the MnGa layer is approximately 446 emu/cm$^3$. As shown in Fig. 2(c), the robust PMA of MnGa is further confirmed by the anomalous Hall effect (AHE) curve, which closely follows the out-of-plane *M*-*H* loop.

As a WSM, CMG exhibits a large intrinsic SHE arising from the Berry curvature. When a charge current is applied along the *x*-direction, a spin current is generated along the *z*-direction with spin polarization along the *y*-direction. When this spin current is injected into the MnGa layer, it exerts a SOT that drives magnetization switching under the assistance of an in-plane magnetic field $H_x$, as illustrated in Fig. 2(a). The sample was patterned into a Hall-bar device with a width of 10 $\mu$m and a length of 20 $\mu$m. A series of 50 $\mu$s current pulses $I_{pulse}$ was then applied concurrently with $H_x$, and the resulting variations in the anomalous Hall resistance $R_{xy}$ were subsequently measured using a 100 $\mu$A probe current.

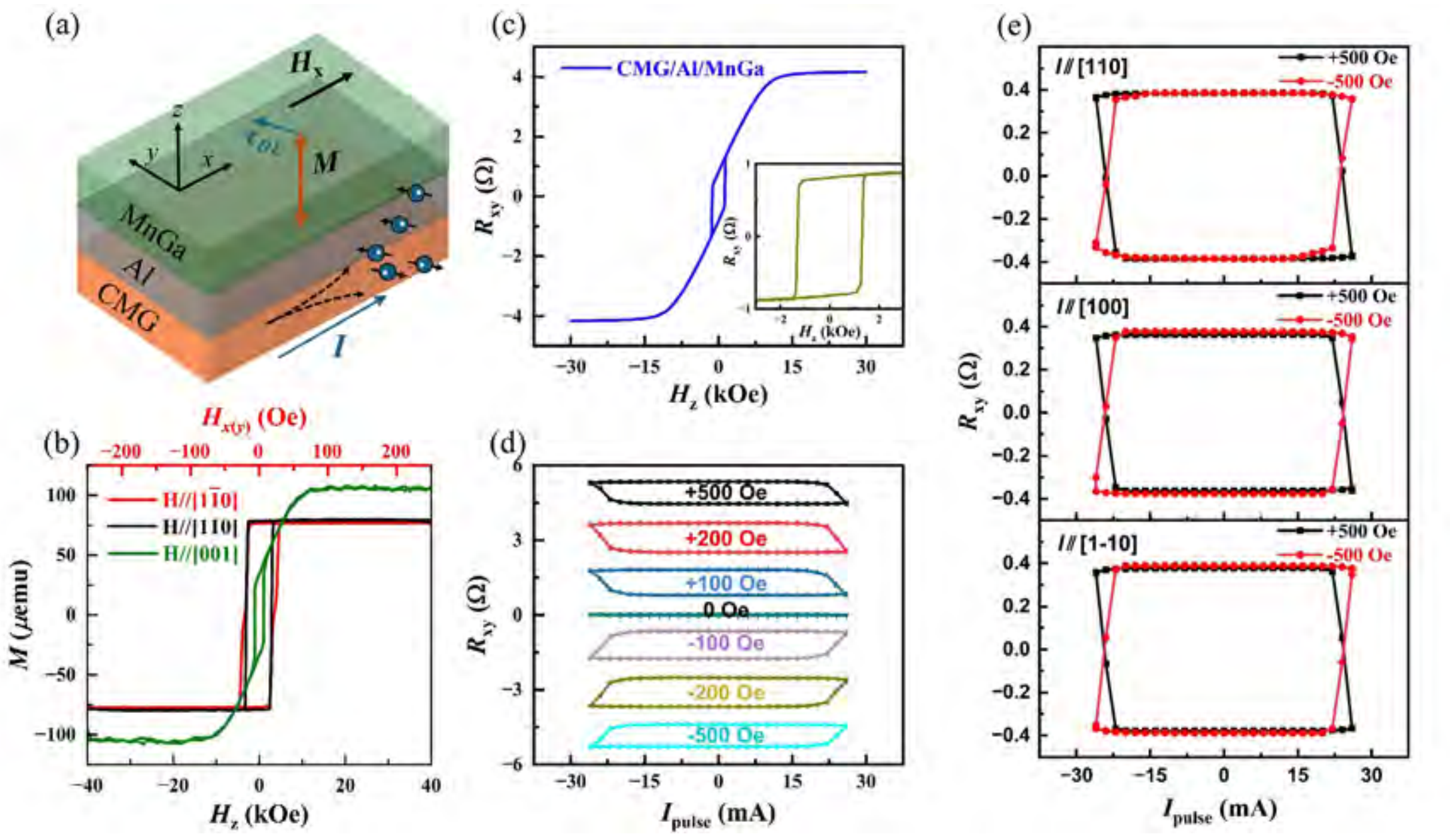

FIG. 2. (a) Schematic illustration of SOT in the CMG/Al/MnGa multilayer. (b) In-plane ([110] and [1-10]) and out-of-plane ([001]) hysteresis loops of the CMG/Al/MnGa multilayer at 300 K. (c) The AHE curve of CMG/Al/MnGa multilayer at 300 K. The inset shows the AHE signal from the MnGa layer after subtracting the CMG contribution. (d) Current-induced magnetization switching curves in CMG/Al/MnGa multilayer under different external field $H_x$. (e) Magnetization switching curves of the CMG/Al/MnGa multilayer when $I_{pulse}$ along different crystalline orientations.

Current-induced magnetization switching was achieved in the decoupled CMG/Al/MnGa multilayer [Fig. 2(d)], with the switching polarity reversing with the sign of $H_x$. The switching ratio reaches 71% under $H_x$=200 Oe, while only a small magnetization switching (~3%) is observed under a zero magnetic field (Supplemental Material S3) [31]. To examine the orientation dependence, we measure the switching curves under $I_{pulse}$ along the [110], [1-10], and [100] orientations. As shown in Fig. 2(e), the switching polarity, amplitude, and critical current density remain consistent along the [110], [1-10], and [100] directions, indicating that the SHE in CMG is essentially isotropic in our system. In contrast to previous reports on CMG/MgO systems where SOT exhibits strong orientation dependence due to epitaxial strain and symmetry breaking [35], CMG films grown on GaAs substrates show negligible anisotropy. The superior lattice matching between CMG and GaAs minimizes epitaxial strain, thereby preserving the magnetic symmetry (Supplemental Material S4) [31].

**Magnetization switching in the coupled system**

To further investigate whether the spin current generated by CMG in CMG/MnGa SAF can induce magnetization switching, we grow the CMG(1.5 nm)/MnGa(3 nm) bilayer. The out-of-plane hysteresis loop of the CMG/MnGa bilayer [Fig. 3(a)] reveals the presence of PMA and a saturation magnetization of 55 emu/cm$^3$, significantly lower than the $M_S$ of either the CMG or the MnGa layers. This reduction provides direct evidence of strong AFM exchange coupling between the CMG and MnGa layers. The AHE curve at 300 K displayed in Fig. 3(b) further confirms the robust PMA in the bilayer. The $R_{AHE}$ of the CMG/MnGa bilayer exhibits a small decrease (increase) around +30 kOe (-30 kOe), which can be attributed to a slight spin-flop transition that occurs in SAF bilayers under strong external magnetic fields [24]. Both the *M-H* and AHE curves suggest that AFM coupling transforms the IMA of CMG into PMA, making it a promising SAF candidate for spintronic applications.

We then investigate the current-driven magnetization switching behavior in the coupled CMG/MnGa bilayer. As shown in Fig. 3(c), current-driven magnetization switching is observed under various $H_x$, with a switching ratio of 69% at -2000 Oe. The switching polarity matches that observed in the CMG/Al/MnGa multilayer, indicating a similar underlying SOT mechanism. Compared with the decoupled CMG/Al/MnGa multilayer, the magnetization switching current density in the coupled system increases from 2.4×10$^7$ A/cm$^2$ to 3.5×10$^7$ A/cm$^2$, while no significant zero-field magnetization switching is observed. The temperature rise during SOT-induced switching, estimated from resistance variation, is approximately 389 K for CMG/Al/MnGa and 447 K for CMG/MnGa, both of which are below the Curie temperatures of the CMG and MnGa (Supplemental Material S5) [31]. Since this bilayer contains no heavy metal layer or other spin current source, the spin current must originate from the CMG layer. For comparison, a CMG/Mn bilayer was also fabricated, in which no spontaneous SOT-driven switching is observed (Supplemental Material S6) [31]. Therefore, we propose that the spin current generated by CMG first switches the magnetization of the MnGa layer. Through strong AFM exchange coupling, this process subsequently drives the reversal of the CMG layer itself, realizing self-induced magnetization switching in the SAF. Compared with SAF/HM-based heterostructures, the CMG/MnGa bilayer eliminates the need for an external spin current source, thereby enhancing the SOT efficiency. Consequently, a lower critical switching current density is observed compared with Co-based Heusler alloy/MnGa/Pt heterostructures [36,37].

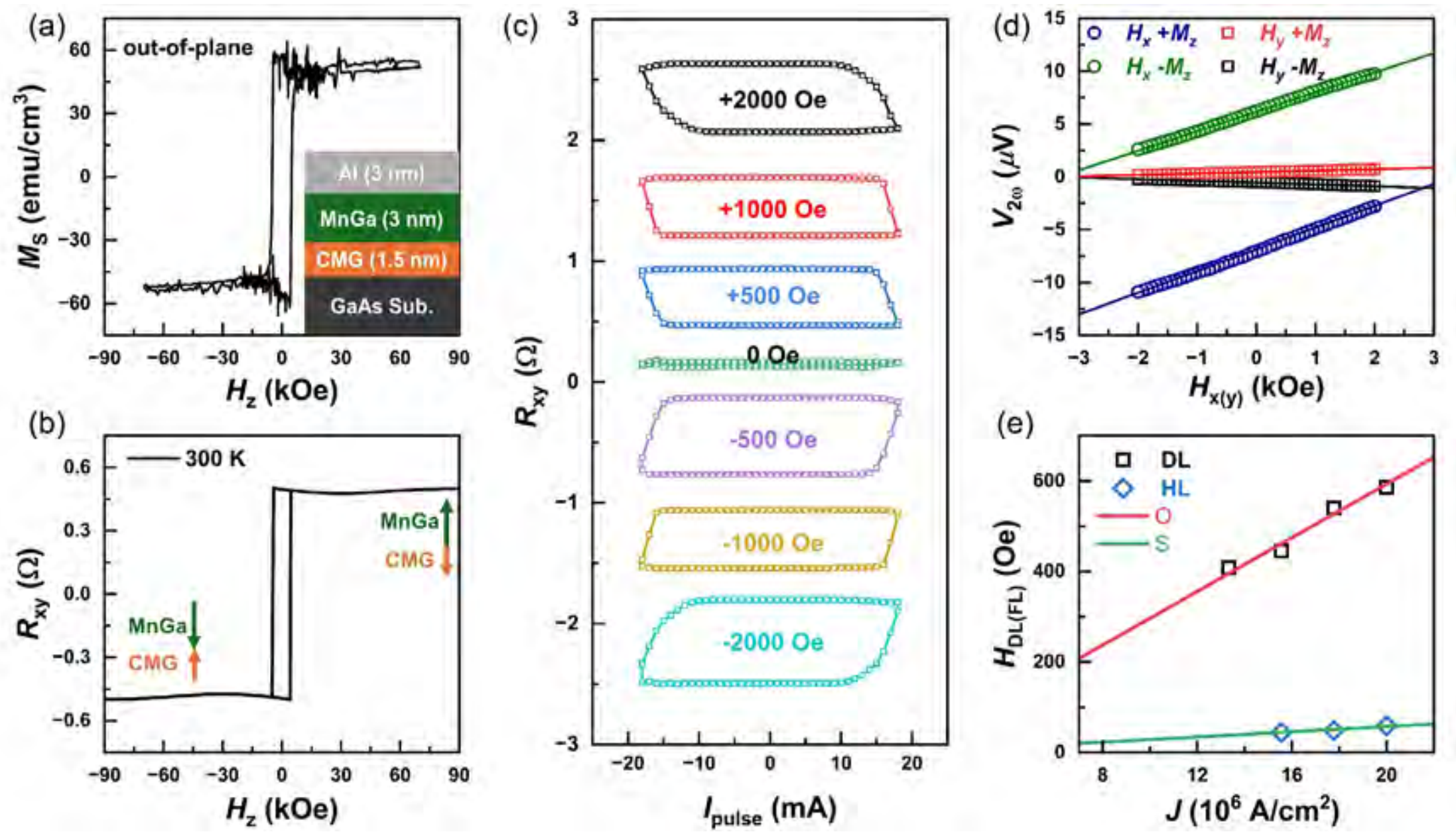

FIG. 3. (a) Out-of-plane *M-H* curve of the MnGa(3 nm)/CMG(1.5 nm) bilayer at 300 K. The inset shows the structure of MnGa/CMG bilayer. (b) AHE curves of the MnGa/CMG bilayer at 300 K. (c) SOT-induced magnetization switching in the MnGa/CMG bilayer under different $H_x$. (d) Variations of second order harmonic signal $V_{2\omega}$ with $H_x$ and $H_y$, respectively. (e) $H_{DL}$ and $H_{FL}$ as a function of applied $J$.

The SOT effective fields in the CMG/MnGa bilayer were quantified by harmonic Hall measurements (Supplemental Material S7) [31]. The second-harmonic Hall voltage $V_{2\omega}^{x(y)}$ shows a linear dependence on the in-plane magnetic field $H_x$ or $H_y$, as shown in Fig. 3(d), allowing the determination of the damping-like ($H_{DL}$) and field-like ($H_{FL}$) effective fields. By linear fitting of the current-dependent data, the effective field efficiencies $\chi_{DL}$and $\chi_{FL}$ are obtained, as shown in Fig. 3(e), defined as

$$\chi_{DL(FL)}=\frac{H_{DL(FL)}}{J}, \tag{1}$$

where $J$ is the current density. The calculated $\chi_{DL}$ and $\chi_{FL}$ are 29.65×10$^{-6}$ Oe·cm$^2$/A and 2.86×10$^{-6}$ Oe·cm$^2$/A, respectively. The effective spin Hall angle of CMG is calculated using

$$\theta_{\mathrm{SH}}=\frac{2eM_s t_{\mathrm{MCA}}}{\hbar}\frac{H_{\mathrm{DL}}}{J}, \quad (2)$$

yielding a value of 0.24. This value is comparable to those reported for other magnetic Weyl semimetals [13]. These results indicate that the SOT in our system is dominated by the DL component, consistent with the intrinsic SHE of CMG.

**Simulation of magnetization switching**

To gain a deeper understanding of the self-induced switching mechanism in the SAF CMG/MnGa bilayer, we perform numerical simulations using a macrospin model for both the coupled and decoupled systems. In the coupled system, the bilayer consists of layer *i* (CMG) and *j* (MnGa), which are coupled via AFM exchange interaction. Based on experimental observations, the SOT acts exclusively on the MnGa layer, while the CMG layer serves as the spin current source. The magnetization dynamics of each layer are described by modified Landau-Lifshitz-Gilbert (LLG) equations [38,39]:

$$\frac{\partial \vec{m}^i}{\partial t} = -\gamma \vec{m}^i \times \left(\vec{H}_{ext} + \vec{H}_K^i + \vec{H}_{ex}^i\right) + \alpha^i \vec{m}^i \times \frac{\partial \vec{m}^i}{\partial t}, \quad (3)$$

$$\frac{\partial \vec{m}^j}{\partial t} = -\gamma \vec{m}^j \times \left(\vec{H}_{ext} + \vec{H}_K^j + \vec{H}_{ex}^j\right) + \alpha^j \vec{m}^j \times \frac{\partial \vec{m}^j}{\partial t} + \gamma H_{DL} \vec{m}^j \times \vec{\sigma}_y \times \vec{m}^j, \quad (4)$$

where $\vec{m}^{i(j)} = \vec{M}^{i(j)}/M_S^{i(j)}$ denotes the normalized magnetization vector of layer *i* (CMG) or *j* (MnGa). The damping-like effective field $H_{\mathrm{DL}}$ is calculated using $\chi_{\mathrm{DL}}= H_{\mathrm{DL}}/J$, and $J$ denotes current density. The key parameters used in the macrospin simulations are summarized in Table 1. Further calculation details are provided in Supplemental Material S8 [31].

The $H_{ex}$ arises from the interfacial AFM exchange coupling and is expressed as:

$$\vec{H}_{ex}^{i(j)} = J_{ex} \frac{\vec{m}^{j(i)}}{M_S^{i(j)} t^{i(j)}}. \quad (5)$$

Equation (5) indicates that $\vec{H}_{ex}^{i(j)} \propto \vec{m}^{j(i)}$, which links the magnetization dynamics of MnGa and CMG. This exchange interaction induces PMA in the CMG layer and enables self-induced magnetization switching. In the decoupled system, only the MnGa layer is considered, with all parameters identical to those in the coupled system except that the AFM exchange coupling term is omitted ($H_{ex}$=0). In this case, the LLG equation accounts solely for the SOT acting on MnGa, allowing us to isolate the effect of spin-current-induced switching without contributions from interlayer exchange interaction.

Table 1. Parameters for macrospin simulation

| **Parameter** | **Definition** | **CMG** | **MnGa** |
|---|---|---|---|
| $\gamma$ | Gyromagnetic ratio | $1.76\times10^{11}$ rad/s·T | $1.76\times10^{11}$ rad/s·T |
| $\alpha$ | Damping factor | 0.005 [19] | 0.015 [40] |
| $t$ | thickness | 1.5 nm | 3 nm |
| $M_S$ | Saturation magnetization | 750 emu/cm$^3$ | 450 emu/cm$^3$ |
| $H_k$ | PMA field | -1000 Oe | 10000 Oe |
| $J_{ex}$ | Exchange coupling energy | -2.5 erg/cm$^2$[23,24] | -2.5 erg/cm$^2$ |
| $\sigma$ | Spin polarization | [0, 1, 0] | [0, 1, 0] |
| $\chi_{DL}$ | SOT efficiency | 0 | $30\times10^{-6}$ Oe cm$^2$/A |

In the coupled system, the initial magnetization directions of MnGa and CMG are set along the [001] and [00-1] orientations, respectively. As illustrated in Figs. 4(a) and 4(b), both layers reverse their magnetization from positive (or negative) to negative (or positive) under the combined influence of the $H_{ext}$ and SOT. The strong AFM exchange coupling ensures that their magnetizations remain antiparallel throughout the entire switching process. Figure 4(c) shows the reversal trajectory of decoupled MnGa. Compared with the coupled MnGa, its trajectory is simpler and smoother. As shown in Figs. 4(d) and 4(e), the coupled MnGa exhibits a faster magnetization switching time than the decoupled MnGa, reflecting high-frequency spin dynamics. The non-identical switching trajectories of CMG and MnGa highlight differences in their dynamic responses, despite their magnetizations remaining antiparallel throughout the process. The CMG layer switches more slowly and exhibits more pronounced oscillations, as it is driven indirectly by the AFM exchange field $H_{ex}$ rather than directly by the SOT. The simulated magnetization switching curves [Fig. 4(f)] demonstrate that CMG and MnGa layers undergo synchronous reversal but require a higher switching current density than decoupled MnGa. In the CMG/MnGa bilayer, magnetization switching must overcome not only the PMA of the MnGa layer but also the additional AFM exchange coupling between CMG and MnGa. This results in a higher effective energy barrier and therefore a larger required switching current density, in good agreement with experimental observations. These simulation results suggest that the SOT generated in CMG first reverses the PMA MnGa layer, which in turn triggers the reversal of CMG through AFM coupling, realizing self-induced magnetization switching in the SAF structure. This mechanism enables self-induced switching in a SAF without requiring an additional spin source.

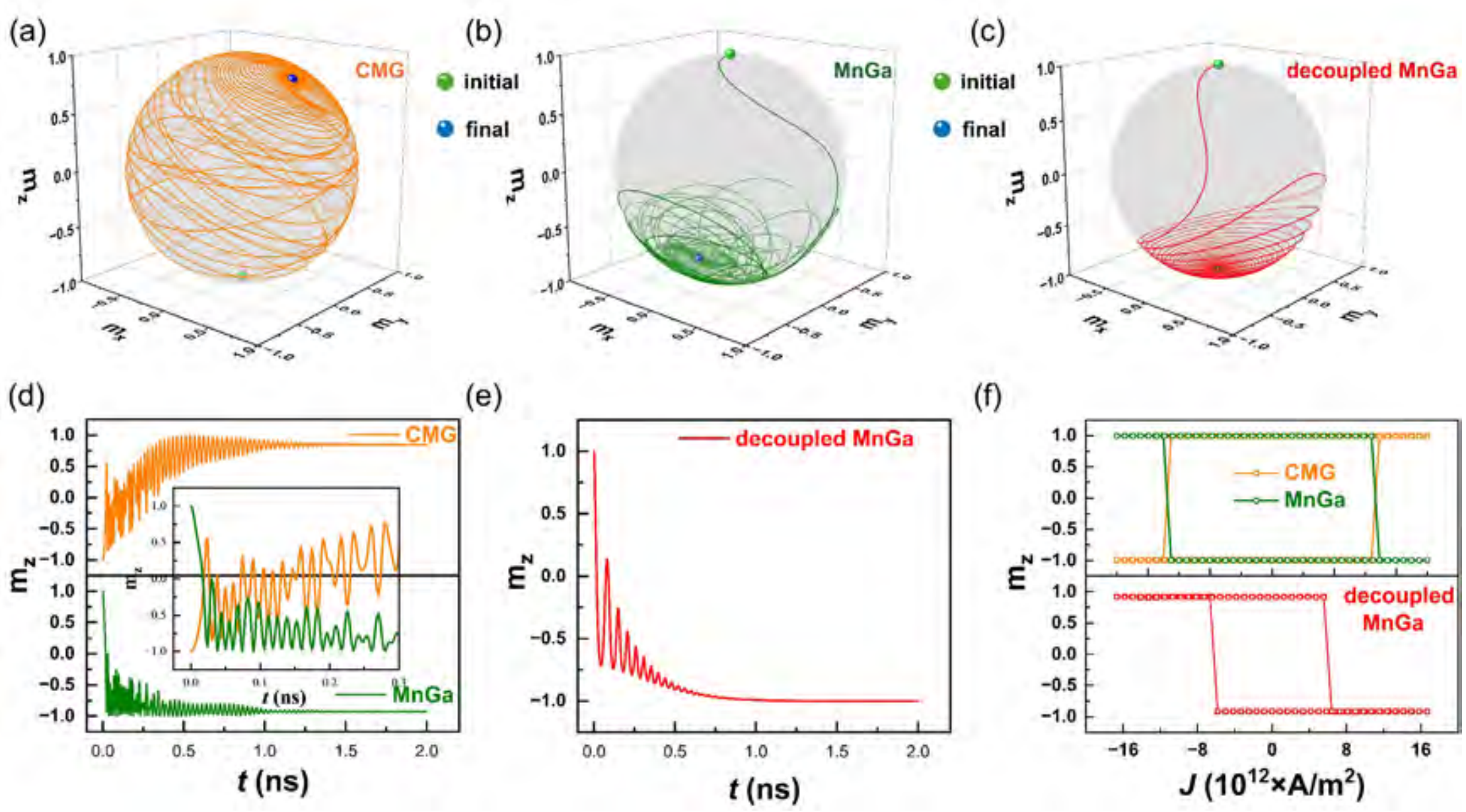

FIG. 4. Trajectories of the magnetization vector *m*. Evolution of the coupled CMG layer from the initial state [00-1] (a), the coupled-MnGa layer from the initial state [001] (b), and the decoupled MnGa layer from the initial state [001] (c). Green and blue dots represent the initial and final state, respectively. (d) Time evolution of the $m_z$ component during SOT-driven switching for the CMG (yellow) and MnGa (green) layers. The inset highlights the details within the first 0.3 ns. (e) Time evolution of the $m_z$ component during SOT-driven switching for the decoupled-MnGa (red) layer. (f) Magnetization switching curves of coupled and decoupled layers under +4000 Oe.

## Conclusion

In conclusion, we have grown high-quality L2$_1$-CMG thin films, demonstrating their potential as efficient spin current sources for magnetization switching. In CMG/Al/MnGa multilayers, we show that spin currents generated by the magnetic WSM CMG can drive magnetization switching of the MnGa layer with strong PMA. In the SAF CMG/MnGa bilayer, self-induced switching is achieved in the CMG/MnGa bilayer without the need for heavy metals, highlighting a new mechanism of SOT-driven magnetization switching. Macrospin simulations confirm that the spin current generated in CMG first switches the MnGa layer, and the strong AFM exchange coupling subsequently drives the reversal of the CMG layer itself, completing the magnetization switching. This self-induced switching mechanism demonstrates that magnetic WSMs can enable autonomous spin control in SAF systems. These findings show that magnetic WSMs can serve as efficient spin current sources for SAF-based devices, paving the way for scalable, energy-efficient AFM spintronic applications.

**ACKNOWLEDGMENTS**

The work was supported by the Research Start-up Funds of Hangzhou International Innovation Institute of Beihang University No. 2024KQ01 and CAS project for Young Scientists in Basic Research under Grant No. YSBR-030. Y.L. acknowledges the support by the French National Research Agency (ANR) SOTspinLED project (no. ANR-22-CE24-0006-01), CHIFTS project (No. ANR-23-CE09-0006-003), and PEPR-SPIN OptoSpinCom project (No. ANR-24-EXSP-0010).

## S1. Origin of AHE in CMG

The anomalous Hall effect (AHE) can originate from intrinsic mechanisms related to Berry curvature and extrinsic mechanisms such as skew scattering and side-jump processes. To investigate the origin of the AHE in $Co_2MnGa$ (CMG), we measured the anomalous Hall conductivity (AHC) $\sigma_{xy}$ at different temperatures, as shown in Figs. S1(a) and (c). Over the measured temperature range, the AHC of CMG exhibits varies only weakly, remaining within the range of approximately 608 to 749 $\Omega^{-1}cm^{-1}$. Such behavior is consistent with an AHE dominated by an intrinsic mechanism, whereas skew scattering contributions are generally expected to show a stronger temperature dependence. For CMG thin films, a high degree of crystalline order is known to be associated with large AHC values, while a reduction in crystalline order leads to a rapid decrease of AHC.

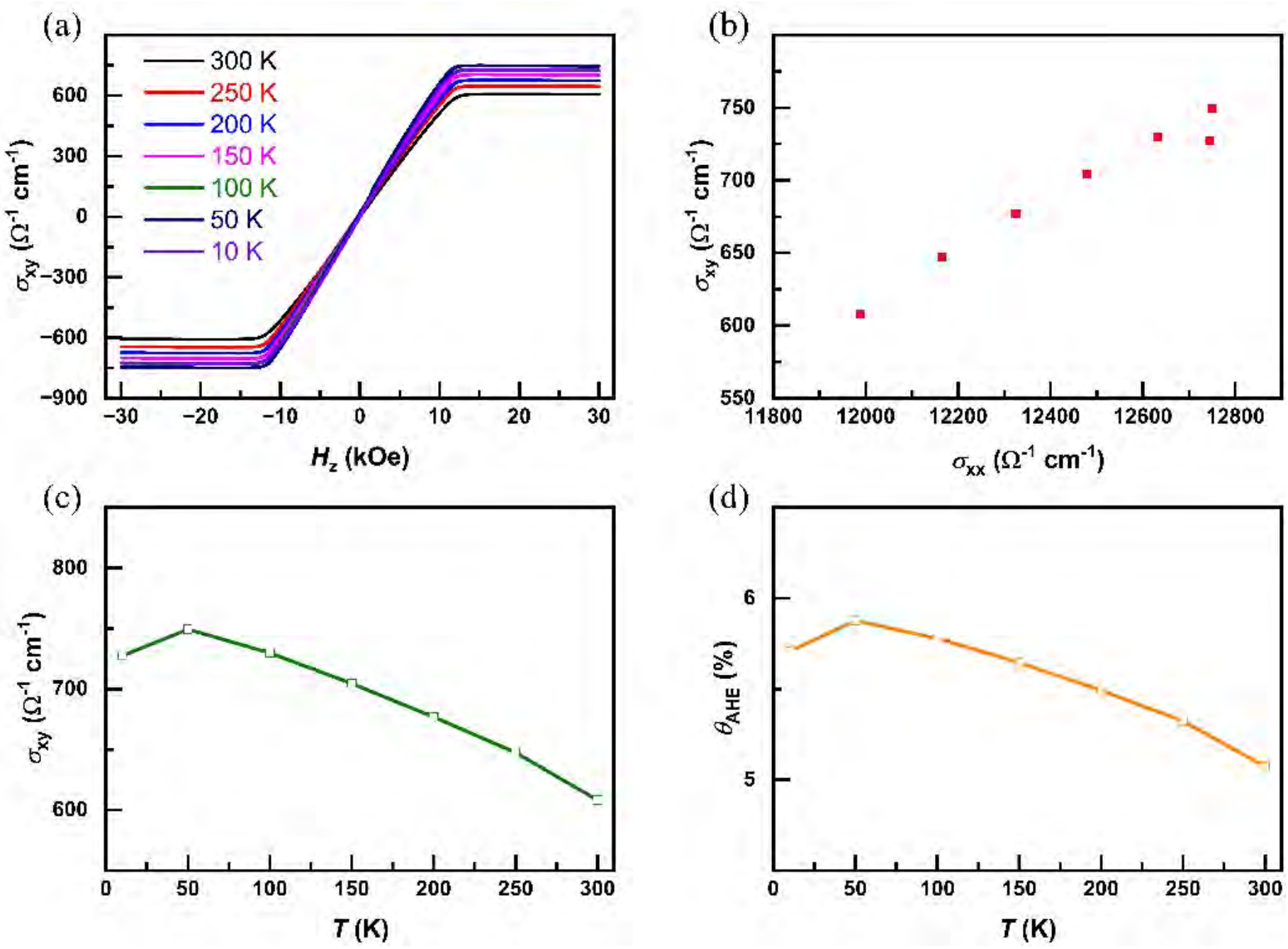

Fig. S1 (a) AHE curve of CMG single layer at different temperature. (b) Anomalous Hall conductivity as a function of longitudinal conductivity CMG single layer. Temperature dependences of (c) anomalous Hall conductivity, and (d) anomalous Hall angle of CMG single layer.

Figure S1(b) shows the AHC as a function of longitudinal conductivity $\sigma_{xx}$ at different temperatures. The $\sigma_{xy}$ remains nearly constant over the measured range of $\sigma_{xx}$, corresponding to the intermediate conductivity regime with $\sigma_{xx}$ on the order of $10^5$ $\Omega^{-1}$

cm$^{-1}$. In this regime, a weak dependence of $\sigma_{xy}$ on $\sigma_{xx}$ is generally regarded as a characteristic feature of predominantly intrinsic AHE contributions. We also evaluated the temperature dependence of the anomalous Hall angle $\theta_{AHE}=\sigma_{xy}/\sigma_{xx}$, shown in Fig. S1(d). At 300 K, $\theta_{AHE}$ is about 5.0% and increases to about 5.8% at 50 K. Such large $\theta_{AHE}$ exceed those typically observed in conventional ferromagnetic thin films with topologically trivial band structures, which usually exhibit values below 3%. The magnitude of $\theta_{AHE}$ and its weak temperature dependence are consistent with a scattering-independent AHE associated with Berry curvature in momentum space, as reported for CMG. Based on the weak temperature dependence of $\sigma_{xy}$, its scaling behavior with $\sigma_{xx}$, and the large $\theta_{AHE}$, the observed AHC is mainly attributed to intrinsic Berry-curvature contributions.

## S2. Magnetoresistance of CMG single layer

We grow a 30 nm-thick CMG single-layer film on a GaAs substrate under the same conditions as for the multilayers. The temperature dependence of the longitudinal resistivity ($\rho_{xx}$) shows a monotonic decrease with decreasing temperature due to the suppression of phonon scattering, displaying a resistivity minimum at approximately 32 K [Fig. S2(a)]. Figure S2(b) presents the longitudinal magnetoresistance (MR) as a function of the in-plane magnetic field $H_x$. The negative MR with a linear field dependence is a notable feature, further supporting the magnetic Weyl semimetal nature of our CMG films.

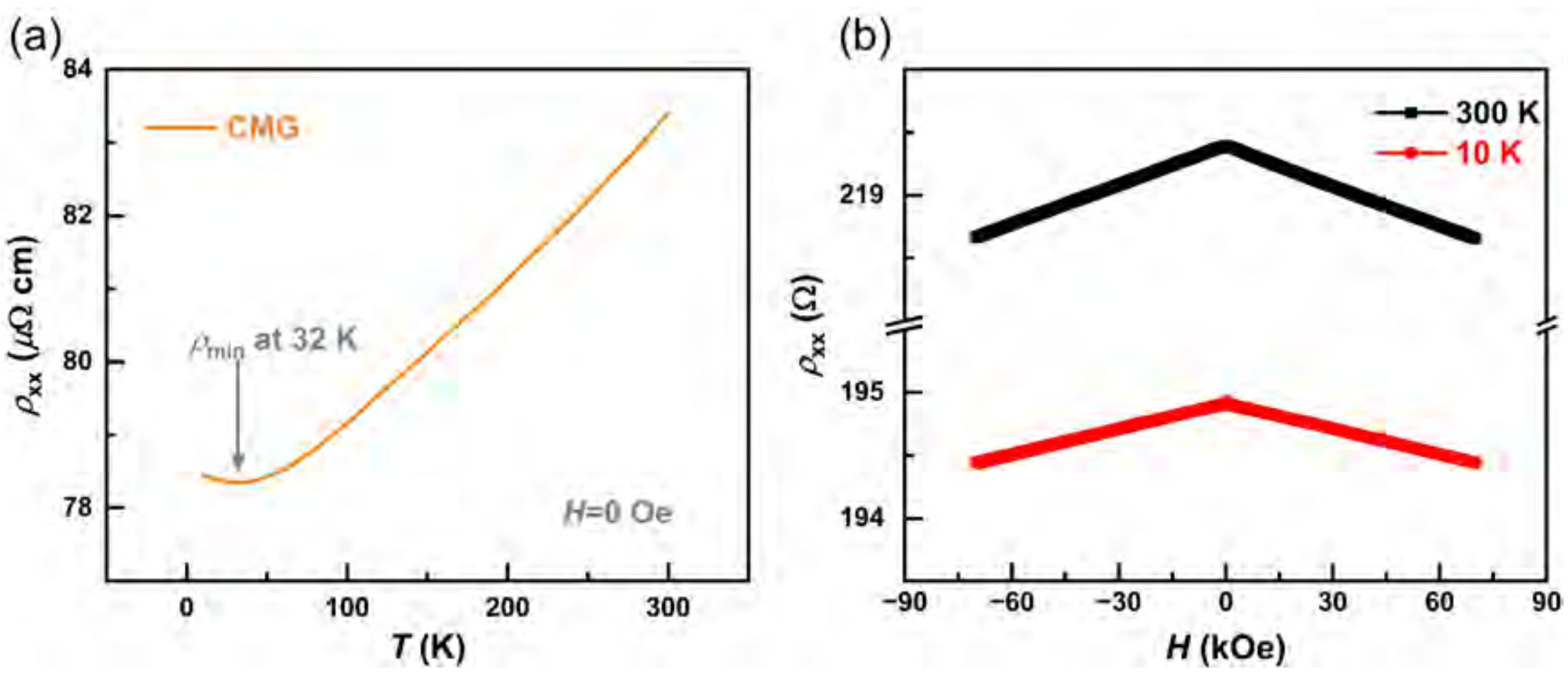

Fig. S2 (a) The longitudinal resistivity $\rho_{xx}$ of CMG single layer as a function of temperature. (b) The longitudinal magnetoresistance $R_{xx}(H)$ curve of CMG single layer, where magnetic is parallel to current.

## S3. The magnetization switching of CMG/Al/MnGa multilayer

We investigate the magnetization switching behavior of CMG/Al/MnGa multilayers under zero external magnetic field. After applying an in-plane field of ±500 Oe and subsequently reducing it to zero, field-free magnetization switching is observed [Fig. S3(a)]. However, the magnetization switching ratio is only about 3%. The chirality of the magnetization switching depends on the history of in-plane magnetic field. When $H_x$ decreases from +500 Oe (or -500 Oe) to 0 Oe, the $R_{xy}$-$I_{pulse}$ curves exhibit clockwise (or counterclockwise) switching. Furthermore, as shown in Fig. S3(b), the application of alternating current pulses of ±26 mA produces highly reproducible magnetization reversals, demonstrating robust switching performance in multilayer.

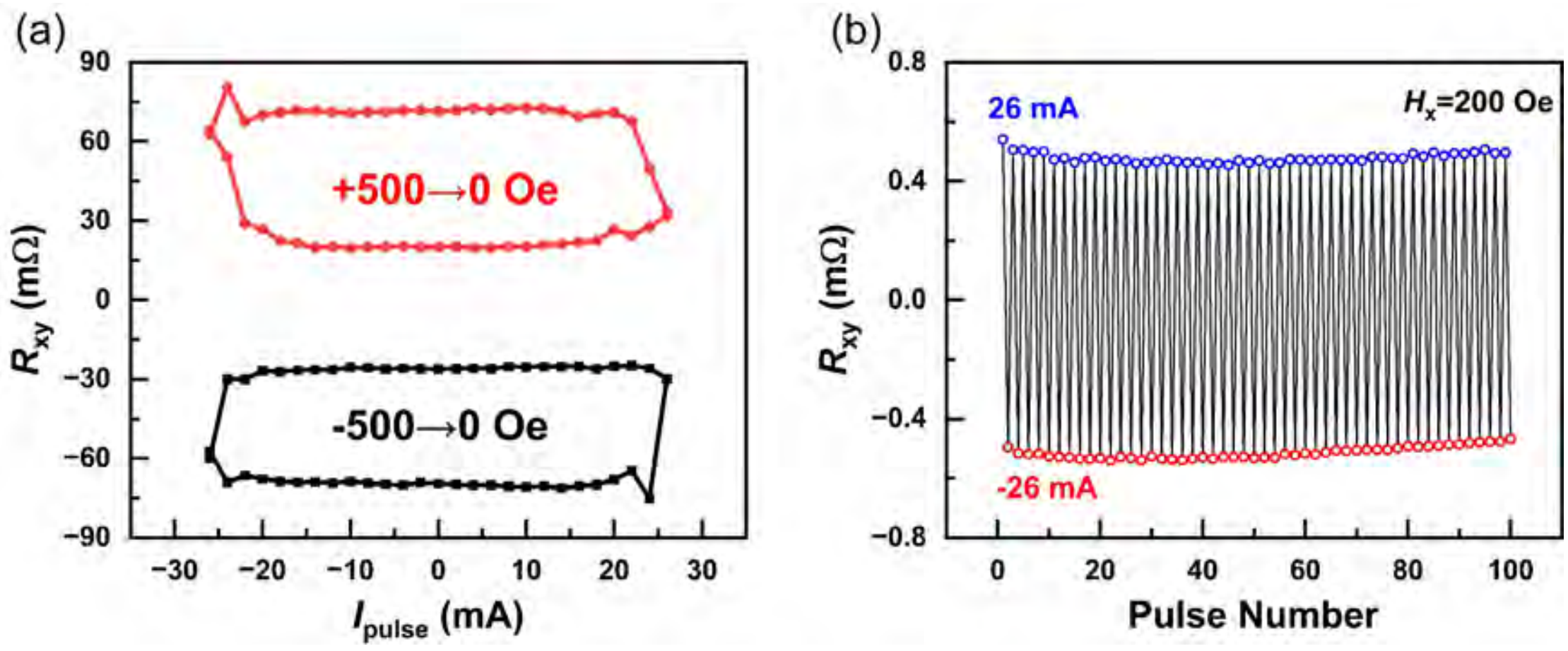

Fig. S3 (a) The field-free magnetization switching loops of MnGa/Al/CMG multilayer after ±500 Oe magnetization. (b) The $R_{xy}$ as the function of pulse number N at 200 Oe. The Hall-bar device was alternately triggered by pulsed current of ±26 mA.

## S4. The in-plane anisotropy of CMG film

We measure the anisotropic magnetoresistance (AMR) of CMG single layer at 1 kOe [Fig. S4(a)]. The AMR $R_{xy}^{AMR}$ is expressed as:

$$R_{xy}^{AMR} = (R_{\parallel} - R_{\perp}) \cos^2 \varphi_M, \tag{1}$$

where $R_{\parallel}$ and $R_{\perp}$ are the resistances when the magnetic moment is parallel and perpendicular to the current, respectively. The $\varphi_M$ is the angle between magnetic moment and the current. Under an applied external magnetic field $H$, the magnetic anisotropy energy ($H_K$) of the film reaches equilibrium with Zeeman energy, which is expressed as:

$$H_K(\varphi_H) = H sin(\varphi_H - \varphi_M), \tag{2}$$

where $\varphi_H$ is the angle between magnetic field and the current. Considering the crystalline symmetry of CMG, both uniaxial anisotropy and fourfold anisotropy are expected, expressed as:

$$H_K(\varphi_H) = \frac{1}{2} H_{ku} \sin 2(\varphi_H - \varphi_{ku}) + \frac{1}{4} H_{k4} \sin 4(\varphi_H - \varphi_{k4}), \tag{3}$$

where $H_{ku}$ and $H_{k4}$ are uniaxial anisotropy and fourfold anisotropy constant, $\varphi_{ku}$ and $\varphi_{k4}$ represent corresponding easy-axis directions, respectively. By fitting the AMR curve at 1 kOe, we find that CMG has a superposition of in-plane fourfold anisotropy ($H_{k4}$=130.6±4 Oe) and uniaxial anisotropy ($H_{ku}$=26.7±2 Oe), as shown in Fig. S4(b). The pronounced fourfold symmetry is consistent with the cubic $L2_1$ crystal structure of CMG, while the weaker uniaxial component likely originates from epitaxial strain induced by the GaAs substrate.

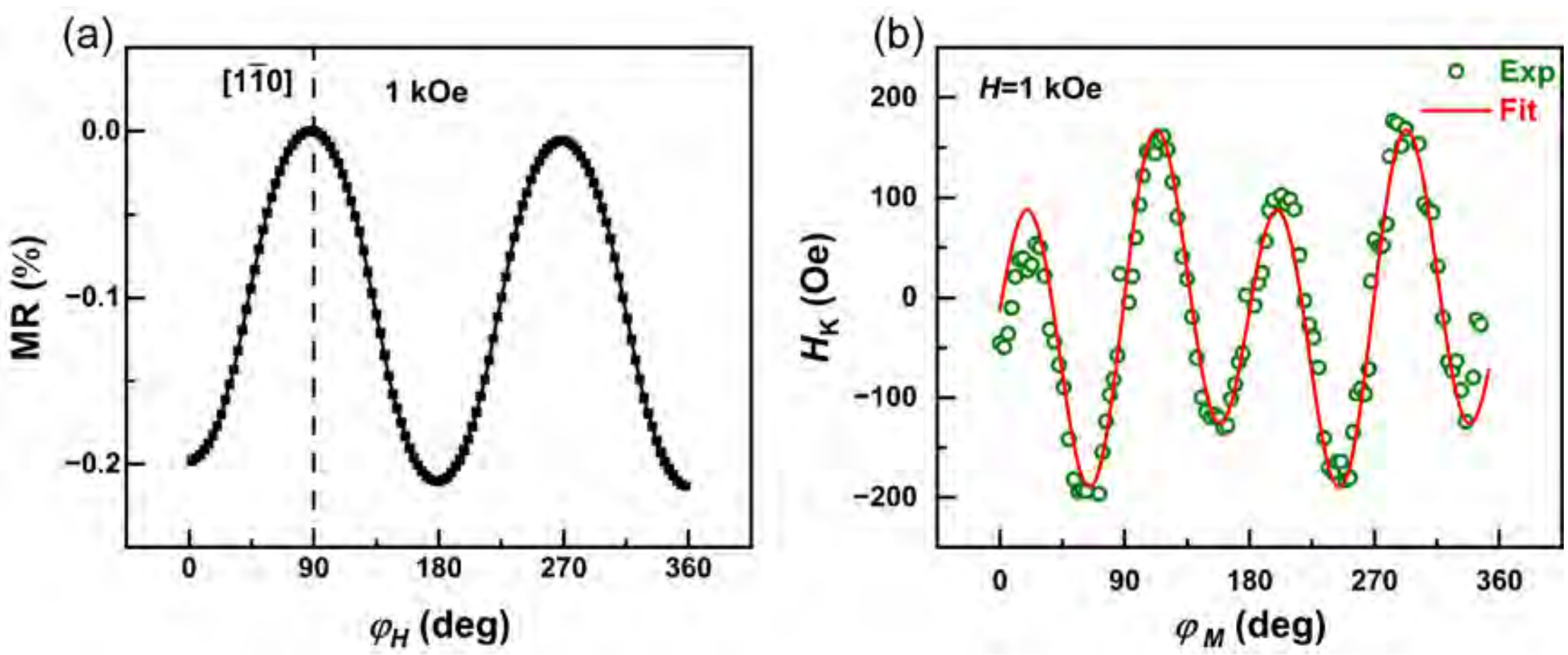

Fig. S4 (a) The AMR curves of CMG single layer at 1 kOe. The dash line is the [1-10] crystalline orientation. (b) The $H_K(\varphi_M)$ curve of CMG single layer at 300 K.

## S5. Joule heating during SOT switching measurement

The perpendicular magnetization switching induced by spin-orbit torque (SOT) usually requires high current densities, which inevitably introduce Joule heating. To evaluate the effect of Joule heating, we measured the $R_{xx}$-$T$ curve and $R_{xx}$-$I_{pulse}$ curves of multilayers, as shown in Figs. S5. Figure S5(a) presents the longitudinal resistance $R_{xx}$ as a function of temperature for the CMG/Al/MnGa Hall bar device. A linear fitting was performed to establish the correspondence between resistance and temperature. Subsequently, $R_{xx}$ was measured as a function of pulsed current $I_{pulse}$, as shown in Fig. S5(b). Owing to instrumental limitations, the minimum pulse width for the measurement was 500 $\mu$s, which is longer than the 50 $\mu$s pulse width employed in the SOT switching experiments. For the CMG/Al/MnGa device, a pulse current of 26 mA yields a resistance of 758.8 Ω, corresponding to a device temperature of approximately 389 K. For the CMG/MnGa device, direct measurements at higher current levels were not possible, because irreversible device degradation occurred when the pulse current exceeded 15 mA. Therefore, the temperature rise was estimated by fitting the measured data using the relationship $T \propto R_{xx} \propto I^2$, as shown in Fig. S5(d). Based on this analysis, a resistance of 1582 Ω at 18 mA corresponds to the temperature of approximately 447 K. These temperatures remain lower than the Curie temperatures of both CMG and MnGa layers. Therefore, Joule heating is unlikely to be the dominant mechanism responsible for the observed magnetization switching.

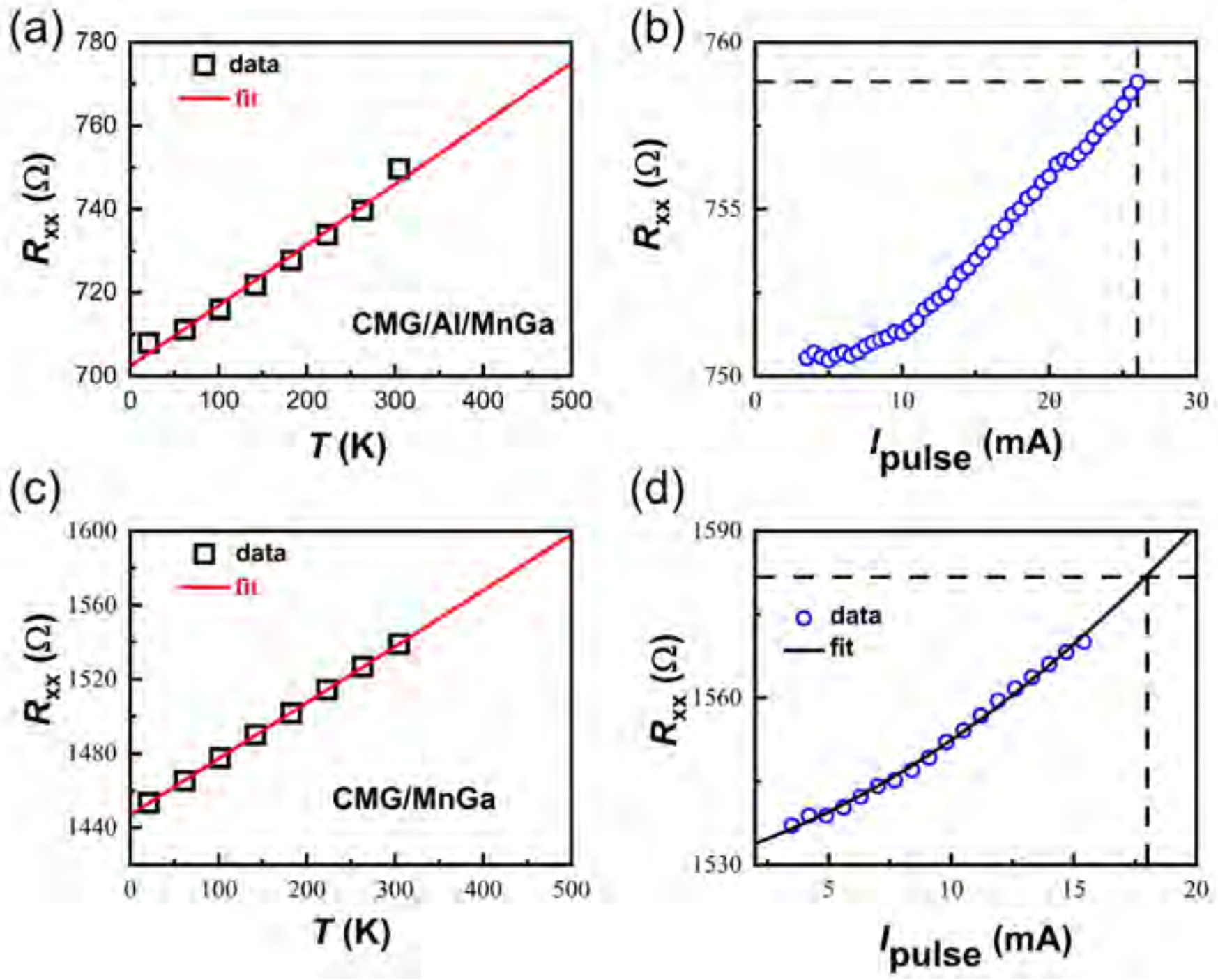

Fig. S5 *R*-*T* curves for the CMG/Al/MnGa (a), and CMG/MnGa (c). The black dots are experimental data, and the red line is the result of linear fitness. The variation of device resistance with different amplitude pulse currents for the CMG/Al/MnGa (b), and CMG/MnGa (d).

## S6. SOT switching in CMG/Mn bilayer

We have investigated whether the CMG layer itself can have current-induced self-SOT switching. Since a single CMG layer is in-plane magnetic anisotropy (IMA), we fabricated a CMG(1.5 nm)/Mn(2 nm) bilayer, where the AFM Mn layer induces perpendicular magnetic anisotropy (PMA) in CMG through interfacial coupling, as shown in Fig. S6(a). However, no magnetization switching was observed before the current exceeded the device breakdown threshold, as shown in Fig. S6(b).

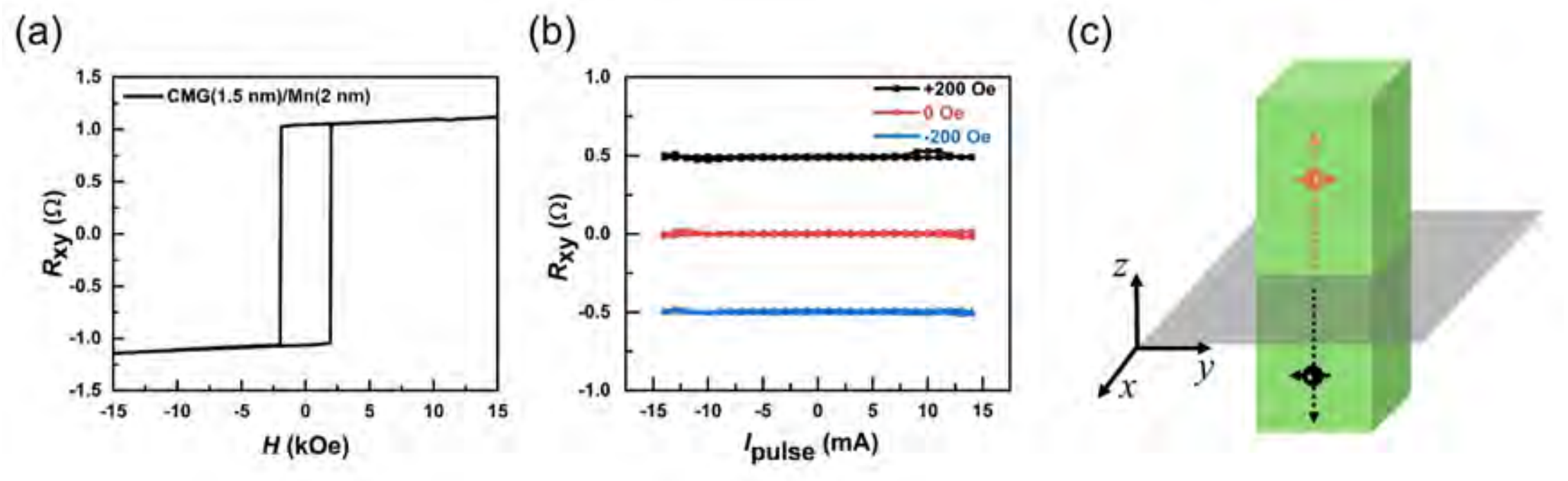

Fig. S6 (a) The AHE curve of CMG/Mn bilayer. (b) The current-induced magnetic switching curves of CMG/Mn bilayer. (c) The spin current in a magnetic single layer with $\mathcal{M}_z$ symmetry.

In CMG, the intrinsic SHE generates a spin current $J_z^y = \frac{1}{2}\{v_z, \sigma_y\} = \frac{1}{2}(v_z\sigma_y + \sigma_y v_z)$, where $v$ and $\sigma$ denote the velocity and spin polarization operators, respectively. However, such a spin current cannot produce a net SOT due to symmetry constraints. For a system possessing mirror symmetry with respect to the *xy* plane [Fig. R1(c)], the mirror operator $\mathcal{M}_z$ transforms the spin current $J_z^y$ as:

$$\mathcal{M}_z J_z^y = \frac{1}{2}(-v_z\sigma_{-y} - \sigma_{-y}v_z) = -J_z^{-y}. \quad (1)$$

This transformation indicates that spin currents with opposite polarization and reversed flow direction are symmetry related. As a result, their contributions to SOT cancel, preventing deterministic magnetization switching. Therefore, CMG does not exhibit SOT-induced magnetization switching. Combined with the absence of SOT-induced switching in the CMG/Mn samples, these results indicate that the switching does not originate from self-switching of the CMG layer. Instead, the adjacent MnGa layer reverses first and subsequently drive the reversal of the CMG layer.

## S7. Harmonic Hall measurement

We have performed harmonic Hall measurements on the CMG/MnGa bilayer to quantitatively determine the SOT effective fields, as shown in Fig. S7(a). By applying an alternating current along the x- or y-axis of the Hall-bar devices, we measured the first- and second-harmonic Hall voltages as functions of in-plane magnetic fields, which allowed us to extract both the damping-like ($H_{DL}$) and field-like ($H_{FL}$) effective fields. The first-harmonic Hall voltage $V_{1\omega}$ signal exhibits a parabolic dependence on the in-plane magnetic field in the low field range. The $V_{2\omega}^{x(y)}$ displays a linear dependence on the in-plane magnetic field $H_x$ or $H_y$ [Fig. S7(b)]. The $H_{DL}$($H_{FL}$) can be extracted using the following relation:

$$H_{DL(FL)} = -2\frac{\partial V_{2\omega}^{x(y)}}{\partial H_{x(y)}} \Big/ \frac{\partial^2 V_{\omega}^{x(y)}}{\partial H_{x(y)}^2}. \tag{4}$$

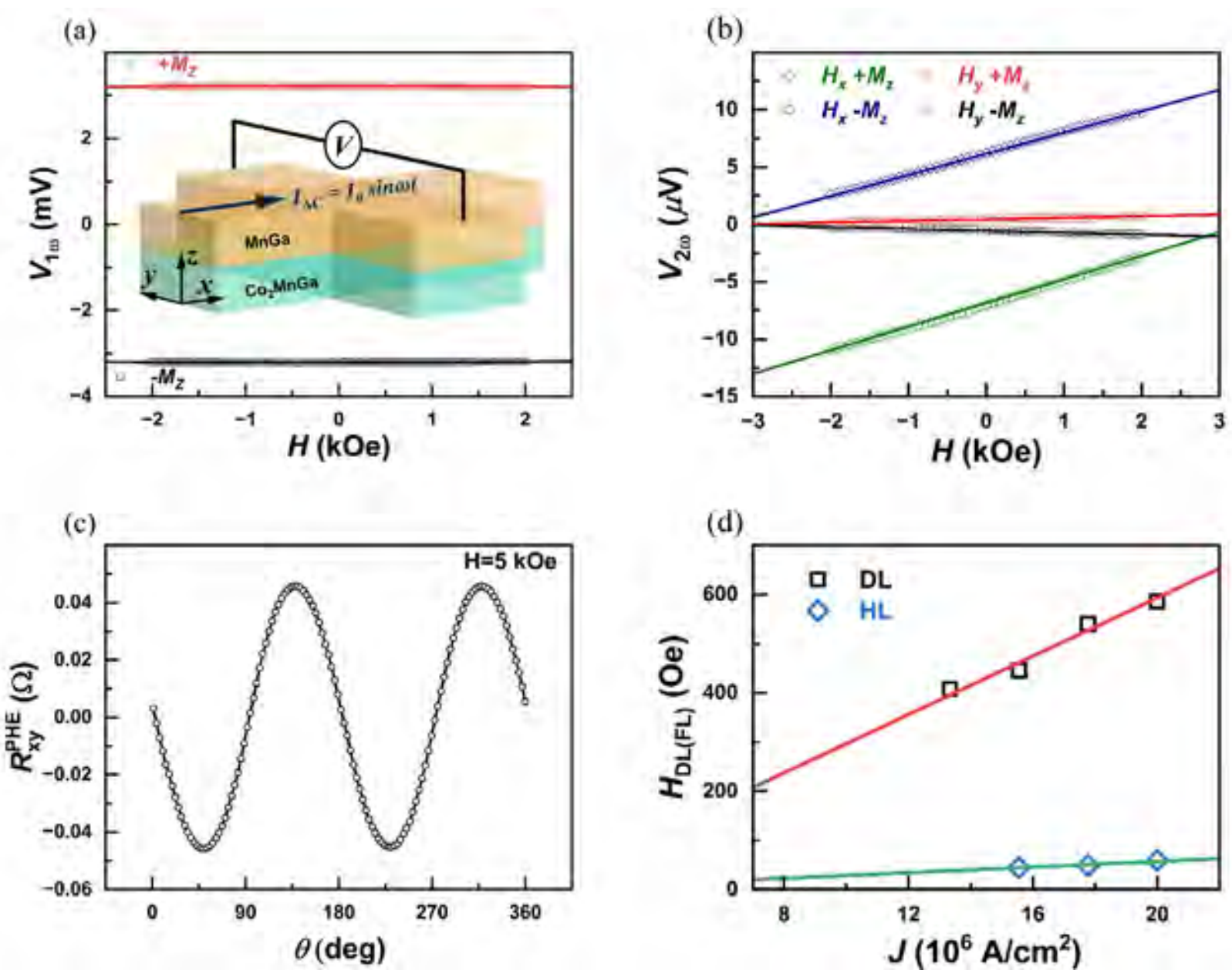

Fig. S7 (a) Variation of first order harmonic signal $V_{1\omega}$ with $H_x$. Inset in the schematic diagram of HHVR measurement. (b) Variations of second order harmonic signal $V_{2\omega}$ with $H_x$ and $H_y$, respectively. (c) The PHE curve of CMG/MnGa bilayer under 5 kOe. (d) $H_{DL}$ and $H_{FL}$ as a function of applied $J$.

The contribution from the PHE is neglected because it is much smaller than the AHE component [Fig. S7(c)]. We considered thermoelectric contributions, particularly the anomalous Nernst effect (ANE). Within the low-field range used for the analysis, the out-of-plane magnetization along remains nearly unchanged. Therefore, the ANE contribution is expected to be approximately field independent and mainly manifests itself as a constant offset in $V_{2\omega}$. As a result,

$$\partial V_{2\omega}^{ANE} / \partial H_x \approx 0. \tag{5}$$

Therefore, its influence on the extracted DL effective field is expected to be negligible.

By linear fitting of the field dependent data, the DL and FL effective field efficiency $\chi_{DL(FL)}$ is obtained as:

$$\chi_{DL(FL)} = \frac{H_{DL(FL)}}{J}, \quad (6)$$

where $J$ is current density. The calculated $\chi_{DL}$ and $\chi_{FL}$i are 29.65×10$^{-6}$ Oe·cm$^2$/A and 2.86×10$^{-6}$ Oe·cm$^2$/A, respectively. The results indicate that the SOT in our system is dominated by the DL component, consistent with the intrinsic spin Hall effect of CMG.

## S8. Macrospin simulation

The magnetization dynamics of the CMG/MnGa bilayer are described by the coupled Landau-Lifshitz-Gilbert (LLG) equations:

$$\frac{\partial \vec{m}^i}{\partial t} = -\gamma \vec{m}^i \times \left(\vec{H}_{ext} + \vec{H}_K^i + \vec{H}_{ex}^i\right) + \alpha^i \vec{m}^i \times \frac{\partial \vec{m}^i}{\partial t}, \quad (7)$$

$$\frac{\partial \vec{m}^j}{\partial t} = -\gamma \vec{m}^j \times \left(\vec{H}_{ext} + \vec{H}_K^j + \vec{H}_{ex}^j\right) + \alpha^j \vec{m}^j \times \frac{\partial \vec{m}^j}{\partial t} + \gamma H_{DL} \vec{m}^j \times \vec{\sigma}_y \times \vec{m}^j, \quad (8)$$

where $\gamma$ is the gyromagnetic ratio, and $\sigma_y$ represents a spin polarization along the $y$-direction. The external magnetic field $H_{ext}$ is applied along the $x$-direction. From experimental measurements, the $M_S$ of CMG and MnGa are 750 emu/cm$^3$ and 450 emu/cm$^3$, respectively, with corresponding layer thicknesses of $t_{CMG}$=1.5 nm and $t_{MnGa}$=3 nm. The Gilbert damping constants are assumed to be $\alpha_{CMG}$=0.005 and $\alpha_{MnGa}$=0.015, , indicating that CMG exhibits lower damping than MnGa. For MnGa, the perpendicular magnetic anisotropy field ($H_k$) is set to 10 kOe, while CMG possesses no intrinsic PMA but instead exhibits in-plane magnetic anisotropy (IMA) originating from the demagnetizing field, for which $H_k$ is set to -1 kOe. Through AFM exchange coupling, CMG can nevertheless maintain an effective PMA. The interfacial exchange constant ($J_{ex}$) is negative, indicating AFM coupling between the two layers. The DL torque efficiency ($\chi_{DL}$) is finite only in the MnGa layer, while the FL torque efficiency ($\chi_{FL}$) is negligible and thus set to zero. The initial magnetization directions are set to (0, 0, -1) for CMG and (0, 0, 1) for MnGa.

The simulation results are shown in Fig. S8. Under an applied magnetic field of +4000 Oe, magnetization reversal does not occur when the current density is below the critical current density ($J_c$). Once the current density exceeds $J_c$, both CMG and MnGa layers switch simultaneously, the results presented in the main text (Fig. 4).

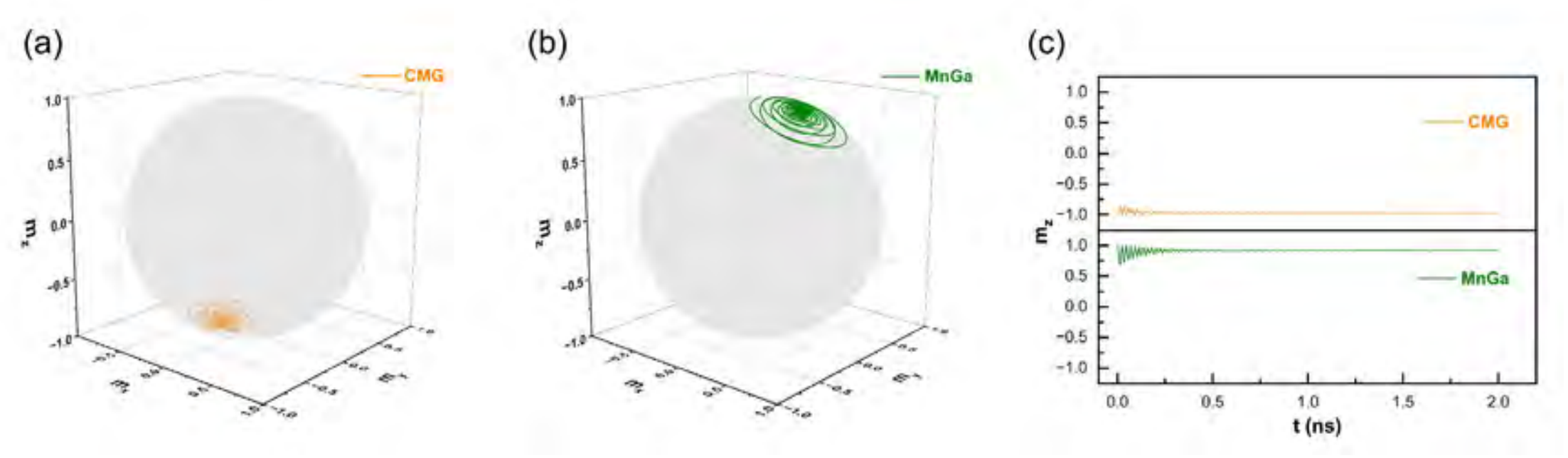

Fig. S8 Trajectories of the magnetization vector *m* of the CMG layer (a), and MnGa layer (b), while $J < J_c$. (c) Time evolution of the $m_z$ component during SOT-driven switching for the CMG (yellow) and MnGa (green) layers.